\documentclass[11pt,a4paper]{article}
\usepackage{hyperref} 

\usepackage[latin9]{inputenc}
\usepackage{float}
\usepackage{rotfloat}
\usepackage{multirow}
\usepackage{amsmath}
\usepackage{amssymb}
\usepackage{graphicx}
\usepackage{comment}
\usepackage[authoryear]{natbib}
\usepackage{caption}
\usepackage{xcolor}
\usepackage{booktabs,caption}
\usepackage{enumitem}
\usepackage[flushleft]{threeparttable}
\usepackage{lineno}
\usepackage{arydshln}
\usepackage[affil-it]{authblk}

\usepackage{natbib,har2nat,url}

\makeatletter

\pdfpageheight\paperheight
\pdfpagewidth\paperwidth

\let\SF@@footnote\footnote
\def\footnote{\ifx\protect\@typeset@protect
    \expandafter\SF@@footnote
  \else
    \expandafter\SF@gobble@opt
  \fi
}
\expandafter\def\csname SF@gobble@opt \endcsname{\@ifnextchar[
  \SF@gobble@twobracket
  \@gobble
}
\edef\SF@gobble@opt{\noexpand\protect
  \expandafter\noexpand\csname SF@gobble@opt \endcsname}
\def\SF@gobble@twobracket[#1]#2{}

\usepackage{anysize}
\usepackage{amsfonts}

\marginsize {2.5cm} {2.5cm} {2cm} {2cm}
\newcommand{\blind}{0}

\makeatother

\begin{document}

	
\if0\blind

\title{The expediting effect of monitoring on infrastructural works. A regression-discontinuity approach with multiple assignment variables} 

\author{Giuseppe Francesco Gori, Patrizia Lattarulo, Marco Mariani}
\affil[]{\small{IRPET -- Regional Institute for the Economic Planning of Tuscany}}

\date{}
\maketitle

\begin{abstract}
\noindent Decentralised government levels are often entrusted with the management of public works and required to ensure well-timed infrastructure delivery to their communities. We investigate whether monitoring the activity of local procuring authorities during the execution phase of the works they manage may expedite the infrastructure delivery process. Focussing on an Italian regional law which imposes monitoring by the regional government on "strategic" works carried out by local buyers, we draw causal claims using a regression-discontinuity approach, made unusual by the presence of multiple assignment variables. Estimation is performed through discrete-time survival analysis techniques. Results show that monitoring does expedite infrastructure delivery.
\end{abstract}
\fi

\if1\blind

\title{The expediting effect of monitoring on local infrastructural works. A regression-discontinuity approach} 

\author{}

\date{}
\maketitle

\begin{abstract}
\noindent Decentralised government levels are often entrusted with the management of public works and required to ensure well-timed infrastructure delivery to their communities. We investigate whether monitoring the activity of local procuring authorities during the execution phase of the works they manage may expedite the infrastructure delivery process. Focussing on an Italian regional law which imposes monitoring by the regional government on "strategic" works carried out by other local buyers, we draw causal claims using a regression-discontinuity approach, made unusual by the presence of multiple assignment variables. Estimation is performed through discrete-time survival analysis techniques. Results show that monitoring does expedite infrastructure delivery.
\end{abstract}
\fi

\if2\blind
\title{The expediting effect of monitoring on local infrastructural works. A regression-discontinuity approach} 

\author{Giuseppe Francesco Gori, Patrizia Lattarulo, Marco Mariani}
\affil[]{\small{IRPET -- Regional Institute for the Economic Planning of Tuscany}}
\date{}
\maketitle
\fi

\newpage

\section{Introduction}
 \label{Introduction}

Public infrastructural investment is widely acknowledged as a major driver of growth and competitiveness.
However, the strength of the infrastructure-growth nexus crucially depends on many context factors. In fact, the type, localisation and cost of infrastructures are key factors in order to trigger relevant economic effects both in the short-run and over a longer time horizon \citep{Crescenzi2012, Elburz2017, Vaelilae2020, Bogart2020}. 
It is, moreover, increasingly recognised in the literature that the Governments' ability to understand local needs and invest in pursuing such needs - bearing adequate costs and in a timely manner- is of paramount importance to ensure returns from public resources at the central and local levels \citep{Castells2005, Dabla2012, IMF2014, Acemoglu2015, Brueckner2015, Crescenzi2016}.  Unfortunately, such quality of government cannot be taken for granted, and the impact of infrastructure can be undermined by inefficiencies arising in all phases, from planning to execution. Public works may suffer, during their execution, from cost increases and long durations. This paper focuses on the issue of work durations: public works taking a long time to complete may imply the dissatisfaction of collective needs and also lead to the delivery of manufactures at a time when they no longer serve the purpose they were planned for \citep{Lewis2011}.

Both time and cost inefficiencies have been widely investigated by the public procurement literature. A very established view in this literature sees poor execution performances of public works mainly originating from imperfect regulatory frameworks that may prevent the appropriate enforcement of procurement contracts. To this regard, attention has been extensively devoted to auction formats \citep[e.g.,][]{Dimitri2006, Bajari2009, Bucciol2013, Decarolis2018a, Coviello2018a, Dechiara2020}, to the degree of contract definition \citep[e.g.,][]{Krahmer2011, Dalpaos2013, Lewis2014, Arve2016} and, more recently, to the role of systemic/institutional features such as the "rule of law" \citep{Coviello2018, Estache2018}.
A less beaten track of research looks at the specific role played by the characteristics of the action of the contracting administrations during the entire contract's life cycle.  According to this view, inefficiencies can originate from the lack of know-how and insufficient experience of some procuring authorities, especially small municipalities \citep{Brown2003, Bandiera2009, Guccio2014, Warren2014, Saussier2015, Baldi2017, Gori2017, Decarolis2018b}. 
Therefore, monitoring the action of such procuring authorities may prove useful to guarantee the quality of their action and to ensure economic impact from infrastructures \citep[e.g.,][]{Grigoli2013, Finan2017, Lattarulo2020}. 
To the best of our knowledge, no empirical study to date has tried to understand to what extent monitoring the action of local procuring authorities may improve the situation. This is a gap that the current paper starts to fill, by looking at whether monitoring the action of local governments during the works' execution phase leads to faster infrastructure delivery.
In a public administration perspective, monitoring the action of a procuring authority is aimed at encouraging the proper management of public funds \citep{Finan2017}, including those -- such as the European Regional Development Fund -- that are also devoted to infrastructure investment at the regional level \citep{Farole2011}. 

The opportunity to investigate the effect of monitoring on the execution times of public works managed by local buyers is provided by an Italian regional law, requiring municipalities to perform careful checks on the execution of works they procured out and report to the regional administration, provided these works have certain characteristics . 
In particular, the regional parliament of Tuscany, in central Italy, passed Law 35 in 2011, stating that this monitoring regime must be enforced on all projects that are deemed strategic, because they meet a financial size and benefit from co-financing by the regional government above certain thresholds.  This law was specifically designed to expedite the time performance of local public works.

To estimate the causal effect of monitoring on the execution duration of public works managed by local authorities, we exploit the particular assignment mechanism established by Law 35, and use a sharp regression discontinuity approach. Since works' assignment to monitoring is determined by the simultaneous fulfillment of multiple quantitative criteria, the usual regression discontinuity methodologies  may not apply to our case without appropriate extensions. To the best of our knowledge, only the recent contribution by \citet{Choi2018} extends sharp regression discontinuity designs to situations, like ours, where all assignment variables have to be above certain thresholds to determine treatment receipt. Therefore, we use this approach. However, a further complication in our empirical analysis is due to the fact that, at the end of the observation period, the execution of the works in our dataset may be either completed or still ongoing. This circumstance requires to define a causal effect as the contrast, at the cutoff value of both assignment variables, between  the  hazard of work completion with and without monitoring. Building on the ideas developed by Caliendo et al. (2013) with respect to the single assignment variable context, the estimation of such hazards of completion is accomplished using "local" discrete-time duration models, adapted to our multiple assignment variables setting. 

We find that the causal effect of monitoring on time-to-completion is positive at the threshold values of the two assignment variables. In particular, our results suggest that monitoring may soon provide a boost to the work's execution, but that such effect diminishes later in time. 
 
The rest of the paper is organised as follows. Section 2 describes the monitoring policy under investigation and the data used for the analysis. Section 3 discusses the methodology. Section 4 presents the results of the analysis; and Section 5 is devoted to sensitivity checks. Section 6 concludes.


\section{Monitoring policy and data} \label{Data}
As mentioned in the Introduction, the policy under investigation consists of the monitoring activity exerted by the regional government on particular types of local infrastructural projects managed by lower-level contracting authorities. This kind of regional monitoring is expected to speed up the execution of works by incentivising higher attention and more careful control by such authorities during works' execution. In line with the programme evaluation literature, such policy will be referred to as the treatment \citep{Imbens2009}.

Public procurement in Italy sees the presence of different buyers at different government levels. Local governments, and in particular municipalities among them, account for the highest share of the overall works \citep{Decarolis2015,Gori2017}, which makes the point of this paper particularly relevant. 
In general, the activity that any local procurer should exert during the work's execution consists of checks on the work site, aimed at ascertaining whether the speed and the quality of execution proceeds as scheduled. These checks are usually performed by the technical personnel of the procuring authority. Their success, however, cannot be taken for granted, especially if a procurer wavers in enforcing the contract. 

To expedite public works managed by lower-level procurers, the regional parliament of Tuscany passed law No. 35, in 2011, mandating regional  monitoring for local works that, at the same time, meet a financial size of at least 500,000 Euros and benefit from co-finanancing from the regional government of at least 50\% of their value. The monitoring regime introduced by the law requires contracting authorities to perform thorough checks on the execution schedule, and share them with the regional offices.
In the event of execution slowdowns, the regional offices can thus urge specific actions by the contracting authorities. Only in case the latter are facing particularly complex issues, the former may provide \textit{ad hoc} technical, case-specific assistance.
Law 35 does not require additional red tape. Instead, it creates an environment where attention on public works from peripheral contracting authorities is substantively incentivised. 

We use an administrative dataset collecting detailed information on all public contracts in Italy: the database of Italian public contracts BDNCP (\textit{Banca Dati Nazionale Contratti Pubblici}).
Such dataset reports information on the buyers and on contract characteristics, including auction formats, expected costs and durations. Once the contract's execution is completed, information on actual costs and durations is added.
In order to carry out our analysis on the effects of Tuscany's regional law 35/2011 we consider the section of the BDNCP dataset comprising all public works procedures started by Tuscan municipalities from 2011 to 2017, including 5,559 projects. Of these, 219 are the projects subject to the regional monitoring regime mandated by law 35 (Table \ref{Descr}).

\begin{table}[h!]
	\caption{Descriptive statistics.}
	\label{Descr}
	\centering
	\begin{tabular}{l c c c}
		& Not Monitored & Monitored & Overall \\
		\hline
		Share of regional co-financing  & 14.2\% & 79.9\% & 16.8\% \\
		Total project cost (Euros) & 254,456 & 1,696,169 & 311,263\\
		Actual duration of completed projects (Months) & 8 & 16.9 & 11.9 \\
		Share of right-censored projects & 22\% & 21.9\% & 22\%\\
		\hline
		No. of works under investigation & 5,340 & 219 & 5,559\\
		\hline
	\end{tabular} 
\end{table}

The paper aims at evaluating whether the application of monitoring causes a desirable reduction in the execution times of public works. The actual duration of work executions is known only for project -- the vast majority -- that reach their completion before the end of the observation period, i.e. before the 31st December, 2017. 
Of the remaining projects, we only know that they will reach completion at some unknown time point outside of the observation period. These right-censored projects amount to 22\% of all projects (Table \ref{Descr}). Their presence motivates the usage of survival analysis techniques in the following analysis based on the regression discontinuity design methodology. \\

\section{Methodology} \label{methodology}
\subsection{Notation}
\label{notation}
Let $i =\{1,...,5559\}$ denote the project,  $S_{i}^{(1)}$ its financial size, and $S_{i}^{(2)}$ its share of regional co-financing. Also, let $M_i=\{0,1\}$ be the treatment indicator for project $i$, where  $M_i=0$ stands for no monitoring (control status), and $M_i=1$ stands for monitoring (treatment status). Finally, let $t=\{1,...,T\}$ denote a certain time point after the start of the project's execution.

We adopt the potential outcomes framework. Under the assumption  that the potential
outcomes of one unit do not depend on the treatment status of other units (Rubin, 1986), each project $i$ has two potential outcomes at each $t$: $Y_{it}(0)$ if the project is originally assigned to the control status, and $Y_{it}(1)$ if it is originally assigned to the treatment status. In this application, $Y_{it}(0)$ is an indicator for completion occurring at time $t$ in case project $i$ received no regional monitoring, $Y_{it}(1)$ is an inidicator for completion occurring at time $t$ in case project $i$ received such monitoring.

 In theory, the project-level effect of  monitoring is defined as the contrast -- at each time $t$ -- between the two previous potential outcomes, $Y_{it}(1)-Y_{it}(0)$. In practice, these potential outcomes cannot be observed simultaneously for a same project. Therefore, attention shifts towards estimands based on the contrast of the hazard of completion under different treatment conditions, such as $h_{t}(1)-h_{t}(0)$, i.e. on the differential probability that a work at the cutoff values reaches completion in $t$, given that completion was not reached earlier, if it was originally assigned to the treatment rather than to the control status. Attention also shifts towards the assumptions that are required to interpret the previous contrast as a causal effect, in light of the particular mechanism that determines the assignment of projects to monitoring. 
 
In this study, each project $i$ is assigned to treatment as a deterministic function of its financial size ($S_{i}^{(1)}$) and of its share of regional co-financing ($S_{i}^{(2)}$). In particular, monitoring is mandated if project $i$ simultaneously satisfies the following conditions, $S_{i}^{(1)}\geq c^{(1)}$, with $c^{(1)}=500,000$ euros, and $S_{i}^{(2)}\geq c^{(2)}$, with $c^{(2)}=0.5$. On the other hand, project $i$ receives no regional monitoring in the following three situations: ($S_{i}^{(1)}<c^{(1)},S_{i}^{(2)}\geq c^{(2)}$); ($S_{i}^{(1)}\geq c^{(1)},S_{i}^{(2)}<c^{(2)}$); and ($S_{i}^{(1)}<c^{(1)},S_{i}^{(2)}<c^{(2)}$). It may be useful, for the remainder of this paper, to define two further indicators. The first one, $A_{i}^{(1)}$, takes on the value of one if project $i$ is equal or above the cutoff value in terms of financial size, i.e. if  $S_{i}^{(1)}\geq c^{(1)}$, while it takes on the value of zero otherwise. Similarly, the second indicator, $A_{i}^{(2)}$, takes on the value of one if project $i$ is equal or above the cutoff value in terms of regional co-financing, i.e. if  $S_{i}^{(2)}\geq c^{(2)}$, while it takes on the value of zero otherwise. 

\subsection{A regression discontinuity approach}
\label{identification}
The particular assignment mechanism previously described allows to adopt a sharp regression discontinuity approach.
Regression discontinuity designs are highly popular with applied economists and other social scientists because of the possibility they guarantee to recover causal effects in observational settings without invoking particularly strong assumptions \citep{Imbens2008,Lee2010, Escanciano2017, Choi2017}.

In the traditional econometric approach, the assignment variable is considered a pretreatment covariate and, under relatively mild continuity (also termed smoothness) assumptions, inference relies on some form of extrapolation at a cutoff point. A handful of recent methodological papers have explored the possibility to view regression discontinuty designs as locally randomised experiments in a region around the cutoff, requiring assumptions other than smoothness for identification \citep[e.g.,][]{Mattei2016,Cattaneo2015}.
In this study, we will follow the more established econometric strand of the methodological literature, as it already provides some extensions to cases, like ours, where the assignment variables are more than one. 

Much has been written on regression discontinuity designs in settings where there is a single variable determining treatment assignment, and a single cutoff. Some recent papers have extended such methodology to cases with multiple cutoffs and/or rankings related to such assignment variable \citep[e.g.,][]{Cerqua2014,Cattaneo2016}. Designs with multiple, different assignment variables have also been explored in a number of methodological papers, dealing with the situation where only one of these variables has to be above a certain cutoff to determine treatment assignment \citep[e.g.,][]{Jacob2004, Matsudaira2008, Imbens2011, Papay2011, Reardon2012, Wong2013}. Instead, the paper by \citet{Choi2018} provides methodological guidance that is particularly suited for cases, like the one analysed in this study, where treatment receipt depends on multiple assignment variables simultaneously being greater or equal to certain cutoff values, rather than on only one out of multiple assignment variables being greater or equal to a cutoff value.

However, the approach by \citet{Choi2018}, originally formalised for continuous outcomes, requires to be adapted to cases, such the one in this study, where unit-level outcomes are not so, and the estimand of interest is the contrast -- at each time $t$ -- between the hazard of completion under the treatment and the control conditions in correspondence of the cutoff values $c^{(1)}$ and $c^{(2)}$:
\begin{equation}
\tau_{t}=h_{t}(1)| c^{(1)},c^{(2)}- h_{t}(0)| c^{(1)},c^{(2)}.
\end{equation}

\noindent However, it is worth noting that $h_{t}(0)|c^{(1)},c^{(2)}$ involves  potential outcomes that can be associated to three types of original control statuses: $Y_{it}(0)^{A_{i}^{(1)}=0,A_{i}^{(2)}=1}|c^{(1)},c^{(2)}$; $Y_{it}(0)^{A_{i}^{(1)}=1,A_{i}^{(2)}=0}|c^{(1)},c^{(2)}$; and $Y_{it}(0)^{A_{i}^{(1)}=0,A_{i}^{(2)}=0}|c^{(1)},c^{(2)})$. In a similar fashion, the potential outcome associated under the treatment status can be viewed as $Y_{it}(1)^{A_{i}^{(1)}=1,A_{i}^{(2)}=1}|c^{(1)},c^{(2)}$.

The approach proposed by \citet{Choi2018} allows for the possibility that being above one of the two cutoff values may systematically influence the potential outcome, whatever the position with respect of the other cutoff value. Such influence is viewed as a fixed quantity irrespective of treatment assignment and has to be neutralised. To this end, the estimand of interest at each time $t$ can be written as follows:
\begin{multline}
\label{estimand}
\tau_{t}=h_{t}(1)^{A_{i}^{(1)}=1,A_{i}^{(2)}=1}|c^{(1)},c^{(2)}-h_{t}(0)^{A_{i}^{(1)}=0,A_{i}^{(2)}=0}|c^{(1)},c^{(2)}-\\\left[h_{t}(0)^{A_{i}^{(1)}=1,A_{i}^{(2)}=0}|c^{(1)},c^{(2)}-h_{t}(0)^{A_{i}^{(1)}=0,A_{i}^{(2)}=0}|c^{(1)},c^{(2)}\right]-\\ \left[h_{t}(0)^{A_{i}^{(1)}=0,A_{i}^{(2)}=1}|c^{(1)},c^{(2)}-h_{t}(0)^{A_{i}^{(1)}=0,A_{i}^{(2)}=0}|c^{(1)},c^{(2)}\right].
\end{multline}

As recalled at the beginning of this Section, identification in the regression disconinuity approach usually relies on relatively mild assumptions regarding the continuity, at the cutoff, of the average potential outcomes under the control condition (e.g., \citealt{Imbens2008}). 
In a multiple assignment variable setting like the one investigated in this study, the following continuity assumptions must be invoked to guarantee that the causal effect in (2) is unbiased \citep{Choi2018}:
\begin{itemize}
\item[(a)]: $h_{t}(0)^{A_{i}^{(1)}=0,A_{i}^{(2)}=0}|c^{(1)},c^{(2)})=h_{t}(0)^{A_{i}^{(1)}=1,A_{i}^{(2)}=1}|c^{(1)},c^{(2)})$; 
\item[(b)]: $h_{t}(0)^{A_{i}^{(1)}=1,A_{i}^{(2)}=0}|c^{(1)},c^{(2)})=h_{t}(0)^{A_{i}^{(1)}=1,A_{i}^{(2)}=1}|c^{(1)},c^{(2)})$;  
\item[(c)]: $h_{t}(0)^{A_{i}^{(1)}=0,A_{i}^{(2)}=1}|c^{(1)},c^{(2)})=h_{t}(0)^{A_{i}^{(1)}=1,A_{i}^{(2)}=1}|c^{(1)},c^{(2)})$.  
\end{itemize}
Under assumptions  (a), (b) and (c), the estimand in (2) can be re-written as follows, using observed quantities rather than potential outcomes: 
\begin{multline}
\tau_{t}=\lim_{(S^{(1)},S^{(2)}) \rightarrow (c_{+}^{(1)},c_{+}^{(2)})}h_{t}-
\lim_{(S^{(1)},S^{(2)}) \rightarrow (c_{-}^{(1)},c_{-}^{(2)})}h_{t}-\\ \left[\lim_{(S^{(1)},S^{(2)}) \rightarrow (c_{+}^{(1)},c_{-}^{(2)})}h_{t}-\lim_{(S^{(1)},S^{(2)}) \rightarrow (c_{-}^{(1)},c_{-}^{(2)})}h_{t}\right]-\\ 
\left[\lim_{(S^{(1)},S^{(2)}) \rightarrow (c_{-}^{(1)},c_{+}^{(2)})}h_{t}-\lim_{(S^{(1)},S^{(2)}) \rightarrow (c_{-}^{(1)},c_{-}^{(2)})}h_{t}\right].
\end{multline}
 The previous identification assumptions cannot be directly tested, as they involve quantities -- namely the potential outcome of treated projects had they not been treated -- which can never be observed in the data. However, their plausibility will be extensively assessed in an indirect fashion in Section \ref{robustness}.

\subsection{Estimation approach}
\label{estimation}

The presence of right-censored works' execution spells suggests the use of an approach based on survival analysis techniques. 
Although the works' execution durations are expressed in days in the original dataset, the use of a discrete-time approach ensures gains in terms of flexibility, especially when -- as in the case of this study -- the assumption of proportional hazards is not invoked. This advantage comes at the acceptable price of the hazard being defined on some aggregate time periods, rather than on a daily basis. After a thorough inspection of the works' execution durations in the dataset at hand, the chosen aggregate time periods are as follows: up to 6 months; 6--11 months; 12 months or longer.

Building on \citet{Caliendo2013}, the estimation of the  potential outcomes that are contrasted in $\tau_{t}$ is performed through a discrete-time hazard model, which actually takes the form of a pooled generalised linear model with a logit link \citep{Kalbfleisch2011}. Such model is specified in a way that returns estimates at the boundary point, which is defined -- in this sudy -- by the two cutoff values, and runs on dataset where each work is repeated as many time periods as its execution lasts. Unlike  \citet{Caliendo2013}, the analysis conducted here allows for non-proportional odds, which entails that the size of the estimated discontinuity may vary depending on the specific post-treatment time period, as  the coefficients estimated in the model for the variables $M_i$, $A_i^{(1)}$ and $A_i^{(2)}$ are time-specific.

Let $h_t^{c^{(1)},c^{(2)}}$, $t=\{1,2,3\}$, denote the hazard of completion at time $t$ for a work at the cutoff values $c^{(1)},c^{(2)}$. Also, let $P_{i}^{(t)}$ denote three indicators for each of the three time periods; $Q_{i}^{(k)}$, $k =\{1,... , 4\}$, denote four indicators for the quadrants defined by the combined values of $A_{i}^{(1)}$ and  $A_{i}^{(2)}$; and $j = \{1, 2\}$ index the assignment variables $S^{(1)}$ and $S^{(2)}$. Finally, let $D_{i}^{(j)}=S_i^{(j)}-c^{(j)}$ be the distance separating work $i$ from the cutoff value of the assignment variable $j=\{1,2\}$. Such distance is what makes it possible to interpret the coefficients associated with the remaining variables as boundary point estimates.
Taking the logit of $h_t^{c^{(1)},c^{(2)}}$ as an outcome, the following model can be estimated:
\begin{multline}
\label{logit}
logit(h_t^{c^{(1)},c^{(2)}})=\sum_{t=1}^{3}\beta_{0}^{(t)}P_{i}^{(t)}+\sum_{t=1}^{3}\beta_{1}^{(t)}M_{i}P_{i}^{(t)}+
\sum_{j=1}^{2}\sum_{t=1}^{3}\beta_{2}^{(jt)}A_{i}^{(j)}P_{i}^{(t)}+\sum_{k=1}^{4}\sum_{j=1}^{2}\beta_{3}^{(kj)}Q_{i}^{(k)}D_{i}^{(j)} 
\end{multline}
where all the coefficients have the form of log-odds. Then, such coefficients can be used to predict the potential outcomes at the cutoffs that are involved in the estimation of $\tau_{t}$ (see Equation \ref{estimand}). In particular, $exp(\beta_{0}^{(t)})/[1+exp(\beta_{0}^{(t)})]$ predicts the potential outcome $h_{t}(0)^{A_{i}^{(1)}=0,A_{i}^{(2)}=0}$, while $exp(\beta_{0}^{(t)}+\beta_{1}^{(t)})/[1+exp(\beta_{0}^{(t)}+\beta_{1}^{(t)})]$  directly predicts the quantity given by  $h_{t}(1)^{A_{i}^{(1)}=1,A_{i}^{(2)}=1}-\left[h_{t}(0)^{A_{i}^{(1)}=1,A_{i}^{(2)}=0}-h_{t}(0)^{A_{i}^{(1)}=0,A_{i}^{(2)}=0}\right]- \left[h_{t}(0)^{A_{i}^{(1)}=0,A_{i}^{(2)}=1}-h_{t}(0)^{A_{i}^{(1)}=0,A_{i}^{(2)}=0}\right]$.
Therefore, $\tau_{t}$ is ultimately estimated as:
\begin{equation}
\label{pred_haz}
\tau_{t}=\frac{exp(\beta_{0}^{(t)}+\beta_{1}^{(t)})}{1+exp(\beta_{0}^{(t)}+\beta_{1}^{(t)})}-\frac{exp(\beta_{0}^{(t)})}{1+exp(\beta_{0}^{(t)})} 
\end{equation}
for $t$=\{6- months; 6-11 months; 12+ months\}.

The methodological literature devoted to regression discontinuity designs recommends to trim observations too far away from the cutoff values prior to the start of any local estimation procedure \citep{Imbens2008}. In this study, works are discarded from the analysis if one of the following conditions is verified: $S_{i}^{(1)}<150000$; $S_{i}^{(1)}>1000000$; $S_{i}^{(2)}<0.05$; $S_{i}^{(2)}>0.95$. After such trim, the number of works under investigation falls from 5,559 to 331 (Table \ref{tab:bandwidth}). Then, within the remaining region, a selection procedure of a bandwidth $h$ can be implemented to reach a reasonable compromise between the opposing needs of bias reduction on the one hand, and precision enhancement (through smaller variance) on the other.
 In particular, a very narrow $h$ around the cutoff values is expected to guarantee little bias, at the price of a high variance of estimates. A very wide $h$ is expected to lead to the opposite result. Another advantage from focussing on a subset of units that are relatively close to the cutoffs is that the  estimation can be performed using models with relatively simple specifications, instead of resorting to complex polynomial structures (Imbens and Lemieux, 2009; Gelman and Imbens, 2014). A number of bandwidth selection procedures have been proposed by the literature \citep{Imbens2008, Imbens2012, Calonico2014}. However, all the cited bandwidth selectors were conceived for settings with a single assignment variable. Therefore, their use in the case study under investigation here is not straightforward, as these approaches require to be generalised to the multiple assignment variable setting. To this end, the simple leave-one-out cross-validation procedure proposed by \citet{Imbens2008} is the only one that can be generalised as part as a non-technical paper like this one. In its orginal version, the goal of such cross-validation procedure is to find a bandwidth $h$ guaranteeing that the chosen model specification minimises, at each side of the cutoff value, the root mean squared error (RMSE). The resulting bandwidth may be symmetrical or asymmetrical with respect to the cutoff value, with the choice often depending on the number of observations available at each side \citep{Imbens2008}. In the presence of a continuous outcome variable, the procedure can be extended to the multiple assignment variable setting by comparing, separately in each quadrant, the RMSE from the model associated with all possible pairs of values of the assignment variables, ultimately selecting the pair of values that exhibit the lowest RSME. Since, in this study, the outcome is binary rather than continuous, the RMSE is replaced by the Brier score, which is the mean squared error of the probability forecast. 
 
 \begin{table}[h]
 	\footnotesize
 	\caption{Results of the bandwidth selection process}
 	\label{tab:bandwidth}
 	\centering
 	\scalebox{0.9}{
 		\begin{tabular}{l | c c c | c c c | c c c |}
 			\hline
 			\multirow{2}{*}{Quadrant} & \multicolumn{3}{c}{Original Data}& \multicolumn{3}{c}{After Trimming} & \multicolumn{3}{c}{Optimal Bandwidth} \\ 
 			&  $S_i^{(1)}$/1000 &  $S_i^{(2)}$ & Projects & $S_i^{(1)}$/1000 & $S_i^{(2)}$ & Projects & $S_i^{(1)}$/1000 & $S_i^{(2)}$ & Projects \\
 			\hline
 			$A_i^{(1)}=A_i^{(2)}=1$ & [500; 9595] & [0.5; 1] & 219 & [500; 1000] & [0.5; 0.950] & 65 & [500; 929] & [0.5; 0.846] & 49 \\
 			$A_i^{(1)}=1$, $A_i^{(2)}=0$ & [500; 14100] & [0; 0.5) & 486 & [500; 1000] & [0.05; 0.5) & 48 & [500; 1000] & [0.166; 0.5) & 45 \\
 			$A_i^{(1)}=A_i^{(2)}=0$	& [42; 500)	& [0; 0.5) & 4255 & [150; 500) & [0.05; 0,5) & 108 & [356; 500) & [0.065; 0.5) & 23 \\
 			$A_i^{(1)}=0$, $A_i^{(2)}=1$ & [41; 500) & [0.5; 1] & 599	& [150; 500) & [0.5; 0.95] & 110 & [356; 500) & [0.5; 0.82] & 26 \\
 			\hline
 			Total & & & 5559 & & & 331 & & & 143 \\
 			\hline
 	\end{tabular}}
 \end{table}
\noindent The bandwidths resulting from this procedure are shown in Table \ref{tab:bandwidth}. The number of works falling in the selected bandwidth is 143, corresponding to 340 work*period observations.

\section{Results}\label{Results}
The estimates of all coefficients of the model described in (\ref{logit}), accompanied by their standard errors and 90\% confidence intervals, are shown in Table \ref{tabresult logit}. Standard errors are estimated accounting for observations, i.e. a work $i$ in time period $t$, being clustered at the level of work $i$.

Some of the estimated coefficients, namely $\hat{\beta}_{0}^{(t)}$ and $\hat{\beta}_{1}^{(t)}$, will be involved in predicting the two hazards required -- as in Equation \ref{pred_haz} -- to estimate the causal effect $\tau_{t}$. 
For the moment, it is worth noting that the estimated value of coefficients $\beta_{0}^{(t)}$ increases over time. This fact reveals that the duration dependence of a work at the cutoffs but not meeting any of the two assignment criteria is naturally positive (i.e. the hazard of completion increases when the actual execution time gets longer and longer). Also, from the inspection of the estimated coefficients $\hat{\beta}_{1}^{(t)}$ and their confidence intervals, we get a first intuition that, if such a work had instead been assigned to monitoring, this would have yielded some benefits.

When predicting the hazards in Equation \ref{pred_haz}, the estimated coefficients $\hat{\beta}_{2}^{(jt)}$ and $\beta_{3}^{(kj)}$ will be  neutralised by fixing their associated covariate at the value of zero.
However, before implementing such neutralisation, it may be interesting to look at $\hat{\beta}_{2}^{(jt)}$ and at their confidence intervals. Both reveal whether -- in this specific application -- being above one of the two cutoff values systematically affects the outcome, whatever the position with respect of the other cutoff. From Table \ref{tabresult logit}, it definitely seems that being above each one of the two cutoffs may decrease the hazard of completion of works, especially in the earlier time period(s). If so, it was worthwhile to have these coefficients in the model.

	\begin{table}[h!]\centering
		\begin{threeparttable}
			\caption{Estimated coefficients}
			\label{tabresult logit}
			\begin{tabular}{l l c c c c}
			\hline
			Coeff. & $t$ & Estimate & Standard Error & \multicolumn{2}{c}{90\% C.I.} \\ 
				\hline
				 & 1 [= -6 months] & -0.558 & 0.718  & -1.739	&  0.623 \\
				$\beta_{0}^{(t)}$ & 2 [= 6--12 months] & 1.130 & 0.774  & -0.143	&  2.403 \\
				& 3 [= 12+ months] & 2.113 & 0.912  & 0.613 & 3.613 \\
				\hdashline
				 & 1 [= -6 months] & 7.037&	2.101	&	3.582&	10.492\\
				$\beta_{1}^{(t)}$ & 2 [= 6--12 months] & 4.344&	1.957 &	1.126&	7.563 \\
				 & 3 [= 12+ months] & 3.761&	1.872&		0.681&	6.840 \\
				\hdashline
				 & 1 [= -6 months] & -2.399&	0.992		&-4.030&	-0.767 \\
				$\beta_{2}^{(1t)}$ & 2 [= 6--12 months] & -1.066&	0.936&		-2.605&	0.473 \\
				& 3 [= 12+ months] & -1.769&	1.161&		-3.678	&0.140 \\
				\hdashline	
				 & 1 [= -6 months] & -5.118&	1.874&		-8.200&	-2.036 \\
				$\beta_{2}^{(2t)}$ & 2 [= 6--12 months] & -4.978&	1.744&		-7.846&	-2.109 \\
				& 3 [= 12+ months] & -2.020&	1.434&		-4.378&	0.338 \\		
%
				\hline
			\end{tabular}
			
\begin{tablenotes}
\footnotesize
\item Note: The eight coefficients $\beta_{3}^{(kj)}$ associated with distance to the cutoff values are just technical components of this particular type of local estimation procedure. Therefore, their estimates are not reported in the Table. 
			\end{tablenotes}
		\end{threeparttable}
	\end{table}

	\begin{table}[h!] 
	\centering
	\begin{threeparttable}
		\caption{Estimated causal effects at the cutoffs}
		\label{Predictions}		
		\begin{tabular}{l c c c c c}
			\hline
			Time period & $\hat{\tau}_{t}$ & Standard Error & $p$-value & \multicolumn{2}{c}{90\% C.I.} \\
			\hline
			-6 months &	0.634 &	0.164 &	0.000 &	0.365 &	0.904 \\
			6-12 months & 0.240  &	0.137 &	0.039 &	0.015 &	0.465 \\
			12+ months & 0.105 & 0.083 & 0.102 & -0.031 & 0.241 \\
			\hline
		\end{tabular} 
	\end{threeparttable}
\end{table}
	
Table \ref{Predictions} reports the estimated causal effect $\hat{\tau}_{t}$, in the three time periods chosen, accompanied by cluster-robust standard errors, 90\% confidence intervals, and the $p$-values associated with the null hypothesis that the estimated effect is not positive (right-tailed test). 
The discontinuity in the hazard of completion is estimated at 63.4\% up to 6 months since the work's execution begun,  then it decreases to 24\% in the second execution semester, and finally falls to 10.5\% after one year. The probability that the estimated effect might not be positive is always very low. However, in the last period, such probability is not low enough to satisfy the conventional requirements for statistical significance. These results indicate that a positive effect of monitoring does exist, although it is decreasing over time. In fact, it can be appreciated especially in the early stage of the work's execution, where its 90\% interval prediction ranges from 36.5\% to 90.4\%. Later on, the interval prediction shifts to lower -- but still positive -- values, and  ends up covering zero in the last period.

A likely interpretation of the previous result is that the analysed monitoring scheme may be useful to address relatively minor issues arising during the work's execution, which could notwithstanding lead to some time escalations. Under these circumstances, monitoring may in fact incentivise procurers to do what is needed and  feasible to stick to the original schedule. If, however, the work's execution lasts too long (i.e. one year or longer), a positive effect of monitoring is surrounded by a considerable amount of uncertainty, possibly due to the fact that, by that time, the hazard of completion would be quite high even without monitoring. 
As for projects not reaching completion, it cannot be ruled out that their time escalation is caused by major issues that monitoring alone is not able to solve. Such major issues may include unexpected events of different kinds, often requiring the design and implemetation of \textit{ad hoc} technical and legal solutions.\\
Conducting, in the current three-dimensional setting, the graphical analyses that are usual in two-dimensional regression-discontinuity studies may be not straightforward. However, to graphically explore discontinuities at the cutoffs, we resort, in the Appendix, to the "centering approach" \citep{Wong2013, Cheng2016}, which allows to investigate such discontinuities in the simpler two-dimensional space.

\section{Supplementary Analyses} \label{robustness}
\subsection{Plausibility of the identification assumption}

The possibility to identify and estimate causal effects relies on the continuity assumptions invoked in Section \ref{identification}. Such assumptions state that, in the absence of any intervention, the hazard of completion at the cutoffs would have been the same whatever the quadrant in which works are located. In other words, they rule out that, at least on average, there is any systematic sorting of the observations into the quadrants, which would lead to a discontinuity that cannot be abscribed to the intervention. 

The continuity assumptions are not directly testable. An indirect way to address the issue is to evaluate and test if there is any discontinuity at the cutoffs in the level of covariates, i.e. on predetermined variables that -- by definition -- cannot be affected by the intervention \citep{Imbens2008}. Here, covariates act as pseudo-outcomes on which a pseudo-effect of the intervention has to be estimated. If a  pseudo-effect is found statistically different from zero, this raises suspects on the plausibility of the continuity assumptions that were invoked. 

To link this assessment to the three identification conditions in Section \ref{identification}, we select three different sets of control units depending on their position in terms of the two assignment variables $S^{(1)}, S^{(2)}$.
Then, we estimate the pseudo-effect, each time using the set of treated units joined with an alternative set of controls.
For a continuous pseudo-outcome (covariate), the model that can be used to predict its average, within each alternative treated-control set, is
\begin{equation}
\label{CovariatesModel}
E(Y_i)^{c^{(1)},c^{(2)}} =\gamma_{1}M_{i}+M_{i}\cdot\sum_{j=1}^{2}\gamma_{2}^{(j)}D_{i}^{(j)} +(1-M_{i})\cdot\sum_{j=1}^{2}\gamma_{3}^{(j)}D_{i}^{(j)}
,\end{equation}

where the parameter $\gamma_{1}$ directly quantifies the pseudo-effect. If the pseudo outcome is binary, a \textit{logit} based on a similar linear predictor can be used to forecast the relevant probabilities and check for any discontinuity. The pseudo-outcomes considered are three covariates, drawn from the field literature \citep[e.g.,][]{Bajari2009,Dalpaos2013,Decarolis2015a}: \textit{Expected duration} (number of months of execution set by the contract); \textit{Auction} (= 1 if the tendering procedure is an open or restricted auction, = 0 if the tendering procedure is a negotiation); and \textit{Lowest-bid} (=1 if the award criterion is lowest-bid; = 0 if the award criterion is most economically advantageous tender).
When estimating each of the previous pseudo-effects, a case-specific cross-validation procedure similar to the one described in Section \ref{estimation} is used to determine the optimal bandwidth. 

Table \ref{Covariates_results} reports the estimated pseudo-effects, their standard errors and the $p$-value for the null hypothesis that such pseudo-effects are equal to zero (two-tailed test), which is exactly what we hope for.
Since the null hypthesis is never rejected, there are no reasons to suspect that the identification assumptions stated in Section \ref{identification} are implausible.

\begin{table}[h]
\centering
\begin{threeparttable}
\caption{Covariates as pseudo-outcomes}
\label{Covariates_results}
\small
\centering
\begin{tabular}{l l c c c c }
\hline
Pseudo-outcome & Control set & Pseudo-effect & Standard Error & $p$-value  \\
\hline
Expected duration  (cont.)& $A^{(1)}=1,A^{(2)}=0$ & -4.931 & 6.188  & 0.437 \\ 
 & $A^{(1)}=A^{(2)}=0$ & -10.242 & 7.404 & 0.197 \\
 & $A^{(1)}=0,A^{(2)}=1$  & -9.166 &  6.861 & 0.190 \\
\hdashline
Auction  (1/0) & $A^{(1)}=1,A^{(2)}=0$ & -0.272 & 0.519 & 0.519 \\  
 & $A^{(1)}=A^{(2)}=0$ & -0.291 & 0.516 & 0.573 \\
 & $A^{(1)}=0,A^{(2)}=1$ & -0.294 & 0.534 & 0.581 \\
\hdashline
Lowest bid (1/0) & $A^{(1)}=1,A^{(2)}=0$ & 0.504 & 0.450 & 0.263  \\ 
 & $A^{(1)}=A^{(2)}=0$ & 0.534 & 0.511 & 0.296 \\
 & $A^{(1)}=0,A^{(2)}=1$ & 0.515 & 0.450 & 0.276 \\

\hline
\end{tabular} 
\begin{tablenotes}
\footnotesize
\item Note: All pseudo-effects are estimated using the set of treated units, $A^{(1)}=A^{(2)}=1$, each time joined with an alternative set of control units.
\end{tablenotes}
\end{threeparttable}
\end{table}

%

Additional insights on the plausibility of the identification assumptions may come from searching for discontinuities at the cutoffs in the assignment variables density functions \citep{Imbens2008,Imbens2009}. Such discontinuity is usually interpreted as the result of a manipulation that occurred in the background of the assignment stage, potentially (but not necessarily) leading to sorting. This approach, very fashionable in  regression-discontinuity design applications, makes sense only if one strongly believes that the hypothetical manipulation occurs only in one direction, while it is rather inconclusive otherwise (for further limits of manipulation testing see \citet{Choi2020}). The  methodological literature to date provides solutions applicable only to the standard setting with a single assignment variable \citep{McCrary2008, Cattaneo2018}. The generalisation of such approach to a two assignment variable setting, that is, in a three-dimensional space, is still a challenge ahead for the methodological literature, which can by no means be faced by the current paper. However, to get an intuition on possible density discontinuities at the cutoffs, we may resort, again, to the "centering approach" \citep{Wong2013, Cheng2016}. Such analysis, explained and conducted in the Appendix, provides no support to the idea that manipulation constitutes a serious threat to identification in our analysis.
 
\subsection{Sensitivity to bandwidth choice}
It is recommended by the methodological literature to assess whether the main results of the analysis are to be found  only in the presence of the selected bandwidth. If so, such results may lose some of their credibility \citep{Imbens2008}. Since the selection of an optimal bandwidth is desirable in that it strikes a balance between bias and precision, it may be particularly reasonable to check for the stability of results when the selected bandwidth undergoes small changes that should not alter too much either the amount of accepted bias or the precision of estimates.
The small changes analysed here correspond to one-percent increase or decrease of each of the eight "external" limits of the optimal bandwidth (Table \ref{tab:bandwidth}). It must be stressed that, in this application, a one-percent change in the bandwidth limit very often entails a more than proportionate change in the number of projects falling in the resulting new bandwidth. For instance, a $5\%$ decrease of all the limits of the bandwidth diminishes the number of works by -13\%; while a $5\%$ widening of the limits increases the number of works by 12\%.

\begin{figure}[h!]
	\centering
	\caption{\small Sensitivity to bandwidth choice. Estimated $\tau_t$ and 90\% confidence intervals for different levels of the bandwidth}
	\label{Sensitivity_Graph}
	\includegraphics[width=15cm,height=10cm]{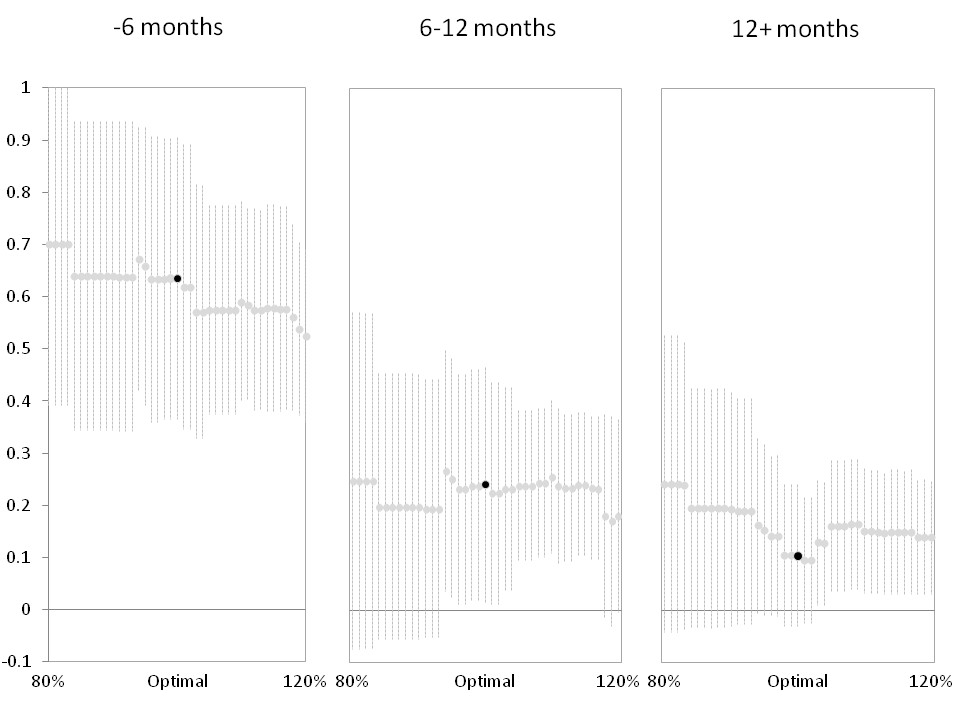}
\end{figure}

Figure \ref{Sensitivity_Graph} reports the estimated effect accompanied by its 90\% confidence intervals, for every variation of the optimal bandwidth in the range $\pm 20\%$. It clearly suggests that the main results of the analysis are not dependent on the single optimal bandwidth that was chosen.

\section{Concluding remarks}
Expediting the final delivery of infrastructural works is a priority of the political agenda of many countries, since economic impact can be undermined by long work durations. Many are the factors that can affect works' completion. Among these, a relevant one  - which has been so far neglected by the empirical literature - regards the opportunity to monitor  local contracting authorities during the execution of the works they commissioned. An adequate level of effort, by local authorities, to stick to the original schedule, could ensure that works proceed at a faster pace, as unexpected hitches can be addressed promptly.

This paper explores the effects of a regional monitoring policy aimed at encouraging such efforts by local authorities, in the attempt to expedite the execution of public works. The policy under investigation is targeted at a subset of public works deemed strategic, whose assignment to monitoring by the regional government is a deterministic function of both their financial size and the share of co-financing by the regional government itself. This peculiar assignment mechanism calls for the adoption of a regression-discontinuity approach based on multiple scores to draw causal claims. Since the outcome of interest is the length of execution, estimation is performed through an appropriately specified, local discrete-time duration model for the hazard of completion of works. 

The results suggest that, at the cutoff values of the two assignment variables, the causal effect of monitoring on time-to-completion is positive, especially during the earlier stages of execution. This kind of result is new in the empirical field literature. Further research is needed to formulate general policy recommendations, including an analysis of the effects of monitoring on final infrastructure costs and, an investigation of its effect away from the cutoffs \citep{Angrist2015}, under assumptions appropriately generalised to the multiple assignment variable setting. However, the first results highlighted here are promising. Increased monitoring could contribute to ensure higher returns from public investment. Given that local infrastructure is often co-financed by regional, national or EU funds, it could also help ensure that strategies formulated at these governance levels are not severely undermined at the moment of their final implementation on the territories.

\newpage

\newpage

\section*{Appendix}
Conducting the usual graphical analyses on the outcome discontinuity at the cutoff, as well as manipulation tests, is particularly demanding with two assignment variables, resulting into a multi-dimensional space. Such a challange is outside the scope of the current paper. To keep things as easy as possible, we may use the "centering approach" \citep{Wong2013, Cheng2016}. Such procedure collapses the two assignment variables $S_{i}^{(1)}$ and $S_{i}^{(2)}$ into a single one, thereby making it possible to conduct the analyses of interest in the two-dimensional space. Since our assignment variables have different measurements units, we first standardise them with respect to their cutoffs values. Thus we obtain the new $V_{i}^{(1)}$ and $V_{i}^{(2)}$, which are comparable and take on the value of zero at the cutoffs $c^{(1)}$ and $c^{(2)}$ respectively. 
To collapse these standardised variables into a single one, we define a new assignment score as the Euclidean distance between each unit's position $(V_{i}^{(1)}, V_{i}^{(2)})$ and the origin $(0,0)$: $Z_{i} = \sqrt{(V_{i}^{(1)}-0)^{2} + (V_{i}^{(2)}-0)^{2}}$. The new assignment score $Z_{i}$ is centered on its cutoff $c^{(Z)}$ and, by construction, such cutoff takes on the value of zero.

\subsection*{Graphical Analysis}

To conduct a two-dimensional space graphical analysis of the discontinuity at the cutoff we focus on the cumulative hazard of completion at the end of the observation period, $H(Z)|T$, which amounts to the probability that a project is completed at some time earlier or equal to time $T$. 
The quantities of interest are as follows:
\begin{itemize}
\item [(A)] $H(Z)^{A_{i}^{(1)} = A_{i}^{(2)} = 1}|T$ represents the cumulative hazard of completion observed for treated projects. Notice that this gross quantity may be affected by treatment but also by the fact of merely being above either cutoff values of the assignemnt variables;
\item [(B)] $H(Z)^{A_{i}^{(1)} = 1,A_{i}^{(2)} = 0}|T$ is the observed cumulative hazard for untreated projects that lie above the cutoff of the financial size assignment variable but below the cutoff of the co-financing assignment variable;
\item [(C)] $H(Z)^{A_{i}^{(1)} = A_{i}^{(2)} = 0}|T$ is the observed cumulative hazard for untreated projects that lie below both cutoff values of the assignemnt variables;
\item [(D)] $H(Z)^{A_{i}^{(1)} = 0,A_{i}^{(2)} = 1}|T$ is the observed cumulative hazard for untreated projects that lie below the cutoff of the financial size assignment variable but above the cutoff of the co-financing assignment variable;
\item [(E)] $\tilde{H}(Z)^{A_{i}^{(1)}=A_{i}^{(2)} = 1}|T$ represents the cumulative hazard of completion for treated projects net of the influence exerted by the fact of merely being above either cutoff values of the assignemnt variables. This quantity is not directly observed in the data. However, building on Section \ref{identification} (equation \ref{estimand} in particular), it can be reconstructed as\footnote{Note that, the net cumulative hazard $\tilde{H}(T)^{A_{i}^{(1)}=A_{i}^{(2)} = 1}$ can be seen as the sum of $H(T)^{A_{i}^{(1)} = A_{i}^{(2)} = 0}$ (quantity (C)) and the discontinuity. Moreover, following \ref{estimand}, the latter discontinuity can be written as $H(T)^{A_{i}^{(1)}=A_{i}^{(2)} = 1} - H(T)^{A_{i}^{(1)} = 0,A_{i}^{(2)} = 1} - H(T)^{A_{i}^{(0)} = 0,A_{i}^{(2)} = 1} + H(T)^{A_{i}^{(1)} = A_{i}^{(2)} = 0}$, yielding the formula in \ref{tilde}.}:
\end{itemize}
\begin{multline}
\label{tilde}
\tilde{H}(Z)^{A_{i}^{(1)}=A_{i}^{(2)} = 1}|T = \\
 (H(Z)^{A_{i}^{(1)} = A_{i}^{(2)} = 1} - H(Z)^{A_{i}^{(1)} = 0,A_{i}^{(2)} = 1} - H(Z)^{A_{i}^{(1)} = 1,A_{i}^{(2)} = 0} +2 \cdot H(Z)^{A_{i}^{(1)} = A_{i}^{(2)} = 0})|T.
\end{multline}

\begin{figure}[h!]
	\centering
	\caption{\small Graphical analysis of the discontinuities in the cumulative hazard of completion at the end of the observation period}
\label{AnalisiGrafica}
\includegraphics[scale=0.8]{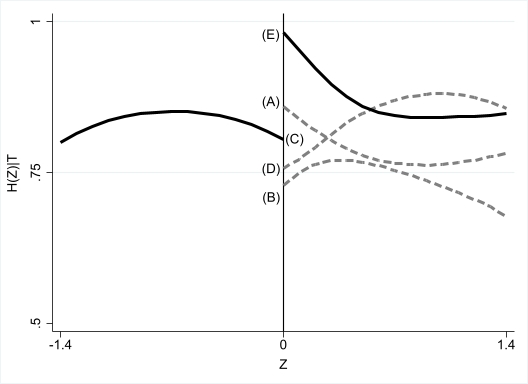}
\caption*{\footnotesize Note: lines (A)-(D) are obtained by plotting the smoothed values from kernel-weighted local polynomial regressions. A triangular kernel polynomial function with degree one is used. Line (E) is then obtained as $(A)-(B)-(D)+2\cdot(C)$, following equation \ref{tilde}.}
\end{figure}

The discontinuity of major graphical interest is the one at the cutoff $c^{(Z)}$, corresponding to the value of zero, of the assignment score $Z$, between the quantities $(E)$ and $(C)$. However, it may be worthwhile to look also at the other discontinuities at $c^{(Z)}$, in order to get insights on how surpassing either thresholds affects the cumulative hazard of completion $H(Z)|T$. 
Figure \ref{AnalisiGrafica} suggests that the discontinuity between $(E)$ and $(C)$ is positive and amounts to $0.18$ points. This graphical result is in line with the main outcomes of the causal analysis conducted in this paper. Moreover, the Figure shows how just surpassing either cutoffs entails downward jump of $H(Z)|T$ which is also in line with the results yielded earlier in this paper.

\subsection*{Manipulation testing}
The goal is to assess the null hypothesis that density $f(.)$ is equal at either side of the cutoff $c^{(Z)}$ of the single assignment score, that is, $H_{0}:\lim_{Z \rightarrow c_{-}^{(Z)}}f(Z) = \lim_{Z \rightarrow c_{+}^{(Z)}}f(Z)$.
We conduct this analysis using the robust bias-corrected statistic and, more in general, the approach developed by \citet{Cattaneo2018}, which in turn builds on \citet{McCrary2008}.\footnote{The anlysis is conducted using the Stata package \textit{rddensity} \citep{Cattaneo2018}.}\\
To make this assessment as informative as possible, we can select different sets of control units depending on their position in terms of the two original assignment variables $S^{(1)}, S^{(2)}$. Test $[1]$ in Table \ref{Manipulation} involves all controls whatever their original position. In such global test we are actually contrasting the density at the cutoff of treated units on one hand, and that of untreated units taken as a whole. Instead, in tests $[2],[3]$ and $[4]$, the density of treated units is contrasted with that of meaningful subsets of controls,
which allows to conduct density discontinuity tests on a more local scale that is logically connected to the three identification conditions in Section \ref{identification}. In order to define these subsets we may recall the notation introduced in Section \ref{notation}, where $A_{i}^{(1)} = 1$ if project $i$ is equal or above the cutoff value in terms of financial size and zero otherwise, and $A_{i}^{(2)} = 1$ if project $i$ is equal or above the cutoff value in terms of regional co-financing and zero otherwise.

{
	\def\onepc{$^{\ast\ast\ast}$} \def\fivepc{$^{\ast\ast}$}
	\def\tenpc{$^{\ast}$}
	\def\legend{\multicolumn{4}{l}{\footnotesize{Significance levels
				:\hspace{1em} $\ast$ : 10\% \hspace{1em}
				$\ast\ast$ : 5\% \hspace{1em} $\ast\ast\ast$ : 1\% \normalsize}}}
	
	\begin{table}[h!] \centering
		\begin{threeparttable}
			\caption{Tests for the null hypothesis of no density discontinuity at cutoff $c^{(Z)}$ under different sets of control units.}
			\label{Manipulation}
			
			\begin{tabular}{l l l c c}
				\footnotesize
				&$Z_{i}<c^{(Z)}$&$Z>c^{(Z)}$&Test statistic&\textit{p}-value \\
				\hline
				$[1]$&All units &$A^{(1)}=1,A^{(2)}=1$ & 1.104 & 0.269 \\
				$[2]$&$A^{(1)}=1,A^{(2)}=0$ & $A^{(1)}=1,A^{(2)}=1$ & 1.506 & 0.132 \\
				$[3]$&$A^{(1)}=0,A^{(2)}=0$ &$A^{(1)}=1,A^{(2)}=1$ & 0.592 & 0.554 \\
				$[4]$&$A^{(1)}=0,A^{(2)}=1$ &$A^{(1)}=1,A^{(2)}=1$ & -1.165 & 0.244 \\
				\hline
			\end{tabular}
			\begin{tablenotes}
				\footnotesize
				\item Note: Robust bias-corrected statistic from local polynomial density estimation developed in \citet{Calonico2019}. A triangular kernel polynomial function of second order is used to construct the density point estimator.
			\end{tablenotes}
		\end{threeparttable}
	\end{table}
}

None of the tests reported in Table \ref{Manipulation} provides enough support against the null hypothesis of no density discontinuity. Therefore, we argue that in our analysis manipulation is unlikely to pose a serious threat to identification.

\bibliographystyle{chicago}
\bibliography{LR35_Final}

\end{document}